\documentclass[aps,prl,twocolumn]{revtex4-1}

 \usepackage[hidelinks]{hyperref}
\usepackage{graphicx,psfrag, subfigure}
\usepackage{bbold}

\usepackage{mathtools}
\usepackage{braket}

\usepackage{array}

\usepackage{amssymb}
\usepackage{color}
\usepackage{eufrak}
\usepackage{graphicx,psfrag, subfigure}
\usepackage{amsmath}
\DeclareMathAlphabet{\mathpzc}{OT1}{pzc}{m}{it}
\usepackage{blkarray}

\usepackage{yfonts}
\usepackage{mathrsfs}
 \usepackage{relsize}
\usepackage{tabularx}

\def\be{\begin{equation}}
\def\ee{\end{equation}}
\def\bea{\begin{eqnarray}}
\def\eea{\end{eqnarray}}

\def\be{\begin{equation}}
\def\ee{\end{equation}}
\def\bea{\begin{eqnarray}}
\def\eea{\end{eqnarray}}

\def\ms{\mathsf}

\def\be{\begin{equation}}
\def\ee{\end{equation}}
\def\bea{\begin{eqnarray}}
\def\eea{\end{eqnarray}}

\def\be{\begin{equation}}
\def\ee{\end{equation}}
\def\bea{\begin{eqnarray}}
\def\eea{\end{eqnarray}}

\newcommand{\td}{\ensuremath{{\text{d}}}}

\usepackage{dsfont}
\usepackage[mathscr]{eucal}

\begin{document}

\title{Vacuum Decay in General Relativity}

\author{Thomas C. Bachlechner}
\affiliation{Department of Physics, University of California San Diego, La Jolla, CA 92093}
\vskip 4pt

\begin{abstract}
We provide a novel, concise and self-contained evaluation of true- and false vacuum decay rates in general relativity. We insist on general covariance and choose observable boundary conditions, which yields the well known false-vacuum decay rate, and a new true-vacuum decay rate that differs significantly from prior work. The rates of true- and false vacuum decays are identical in general relativity. The second variation of the action has a negative mode for all parameters. Our findings imply a new perspective on cosmological initial conditions and the ultimate fate of our universe.

\end{abstract}

\maketitle

{\it Introduction}~---~Consider a gravitational theory that contains two vacua with distinct energy densities. The vacua are stable against perturbative quantum fluctuations, but they can decay via the non-perturbative process of bubble nucleation. In the semiclassical approximation the vacuum decay rate is related to the action $B$ of the bubble formation process that solves the Euclidean equations of motion \cite{Coleman:1977py,Coleman:1980aw}
\be\label{tunnelingpintro}
{ \Gamma}\propto e^{-B/\hbar}~[1+\mathcal{O}(\hbar)]\,.
\ee

We are interested in the thin-wall vacuum decay rate including gravity. This is an old problem that has not received a conclusive answer for all decay channels. The  flaw of  previous arguments lies in the boundary conditions. The traditional approach is based on the Gibbons-Hawking-York boundary action and fixes the global three-geometry of the initial and final states. The three-geometry contains information about how many causally disconnected universes form during the tunneling process and affects the vacuum decay rate. In this formalism the vacuum decay rate  depends on global information that a local observer cannot deduce. This failure is a manifestation of a measure problem, and led to confusion about the rate of false-vacuum bubble nucleation \cite{bgg,Lee:1987qc,fgg,fmp1,fmp2,Aguirre:2005xs,Aguirre:2005nt,Brown:2007sd,BEMtoappear}. We provide a detailed review of this traditional approach in section 3 of \cite{runaways} and show explicitly that the decay rate depends on the (unobservable) number of de Sitter spaces that are nucleated.

In this work we use different, locally observable boundary conditions. Specifically, we demand that there exists a coordinate choice that renders the boundary metric to be of the Schwarzschild-(anti) de Sitter form, with fixed mass $\ms M$. In order for these covariant boundary conditions to yield a well-posed variational problem, we employ the new gravitational action $S_\text{G}$ recently introduced in \cite{tbgravity}. The covariant action $S_\text{G}$ vanishes for all isotropic and stationary spacetimes, and leads to a different vacuum decay rate that no longer depends on causally inaccessible data or coordinate choices.

We find the exponent $B$ of the  true- and false vacuum decay rate in the weak gravity limit
\be\label{weakgintro}
B=\frac{27 \pi ^2  \sigma^4}{2 \left|\Delta \rho\right| ^3}+{\mathcal O}(G)\,,
\ee
where $\sigma$ is the domain wall tension and $\Delta \rho$ is the vacuum energy difference. Our result reproduces the non-gravitational false vacuum decay rate, but also allows for true vacuum decays at the same rate. In contrast to pure field theory, general relativity allows for unimpeded upward vacuum transitions, which implies that the low-entropy initial states required for cosmic inflation are readily attainable and may recur in the future.

\bigskip
{\it Setup}~---~We study the classical and semiclassical properties of a bubble with internal vacuum energy density $\rho_-$, moving in an external region of vacuum energy density $\rho_+$. For simplicity, we assume a thin and spherical shell with surface energy density and tension $\sigma$, separating the bubble interior from the exterior. The full action has gravitational, shell and boundary terms
\be
S=\underbrace{\int_{\cal M}\td ^4 x\, \sqrt{g}\left[ {{\cal R}\over 16\pi G}- \rho(r)\right] }_{S_\text{G}+\dots}-\underbrace{\sigma\int_\text{wall} \td^3{\cal A}}_{-S_\text{Shell}}+S_\text{B}\,, \hspace{-4pt}
\ee
where ${\cal R}$ is the Ricci scalar, ${\cal M}$ is the coordinate region of interest, ${\cal A}$ denotes the domain wall world-volume, and the total derivative terms $S_\text{B}$ render the variational problem well-posed. The ellipses in the gravitational action denote boundary terms that we determine in \cite{tbgravity}. The vacuum energy density is constant both inside and outside the shell located at a radial coordinate $\hat{r}$, 
\be
\rho(r)=\begin{cases}\rho_-\,~~~\text{for~~}r<\hat{r}\\\rho_+\,~~~\text{for~~}r>\hat{r}\end{cases}\,.
\ee
Without loss of generality, we take $\cal M$ to contain the entire trajectory of interest and to be bounded by three-surfaces of constant coordinates of a general metric
\be
t_{\text{i}}\le t\le t_{\text{f}}\,,~~
r_{\text{min}}\le r\le r_{\text{max}}\,.
\ee
In order to avoid breaking general covariance and introducing a measure problem in the form of coordinate dependence in observables, we have to specify covariant boundary conditions at $\partial {\cal M}$, as explained in detail in \cite{tbgravity}. We demand a fixed location $\hat{r}$ of the wall and require that there exist  coordinates that render the metric to be of Schwarzschild-(anti) de Sitter form,
\be
	\label{staticmetric}
		\td s_\pm ^2|_{\partial \cal M} = -A_\pm \td {\ms  T}^2 + A^{-1}_\pm  \td {\ms  R}^2 + {\ms  R}^2 \td\Omega_2^2\, |_{\partial {\cal M}} \,,
\ee
where we defined
\be\label{redshiftfactor}
A_\pm = 1 - \frac{2 G{\ms M_\pm }}{\ms R} - \frac{8\pi G\rho_\pm }{3} \ms R^2 \,,
\ee
subscripts $\pm$ denote the exterior/interior metric, $\ms M_\pm $ is the mass of the solution and $\ms T$ and $\ms R$ denote the Killing time and transverse radius, respectively. Our manifestly covariant boundary conditions define a Dirichlet problem for $\hat{r}$, the mass $\ms M$ and the radius $\ms R$,
\be
\delta{\hat{r}}|_{\partial \cal M}=\delta{\ms M}|_{\partial \cal M}=\delta \ms R|_{\partial \cal M}=0\,. 
\ee

The variables that define our setup are covariant under (smooth) coordinate transformations, so the boundary conditions do not fix any coordinate choices. Correspondingly, the choice of spacetime region ${\cal M}$ is arbitrary and does not affect physical observables. Our boundary conditions are different from those used in prior work on vacuum decay, so we expect to find different observables. In particular, we will find observables that do not depend on the boundary location, e.g. via $r_\text{max}$. This may appear to be an obvious requirement, but it is not obvious. As we demonstrate in \cite{tbgravity}, the commonly used boundary conditions that fix the  boundary metric \cite{York:1972sj,Gibbons:1976ue} yield vacuum decay rates \cite{Coleman:1980aw,Brown:1987dd,Brown:1988kg,Lee:1987qc,fgg} that do depend on the choice of ${\cal M}$  and source much confusion \cite{fmp1,fmp2}. We stress that these non-covariant theories are not  inconsistent, they merely require additional physical input about how coordinates are fixed in nature in order to resolve the associated measure problem.

\bigskip
{\it Action Principle for General Relativity}~---~Let us briefly review the covariant variational principle for isotropic gravity we recently introduced in \cite{tbgravity}. For now we ignore the domain wall and only discuss the vacuum solutions within ${\cal M}$. We can write the most general spherically symmetric metric as \cite{tbgravity}
\be\label{metric}
ds^2=\left(A^{-1}{R'^2}-A{\pi_{\ms M}^2}\right) \left[(\td r+ \ms N_r\td t)^2- \ms N_t^2 \td t^2\right]+ \ms R^2 \td \Omega_2^2\,,
\ee
where $A$ is defined as in (\ref{redshiftfactor}), $\pi_{\ms M}= -\ms T'$  is the momentum conjugate to $\ms M$ and the lapse $\ms N_t$ and shift $\ms N_r$ are non-dynamical variables that impose diffeomorphism invariance. We can easily render the perhaps unfamiliar metric (\ref{metric}) in ADM form \cite{Arnowitt:1962hi,Note5}. \footnotetext[5]{With new functions $R=\ms R$, $\Lambda^2=A^{-1}\ms R'^2-A {\pi_{\ms M}^2}$, $N_t=-\Lambda {\ms N}_t$ and $N_r={\ms N}_r$ we recover the metric in ADM form, $ds^2=-{N_t^2} \td t^2 + \Lambda^2 (\td r + {N_r} \td t)^2 + R^2 \td \Omega_2^2$.} With a suitable choice of boundary terms the action that gives rise to Einstein's equations under the variational principle $\delta S_\text{G}=0$ is given by \cite{tbgravity}
\be\label{gravaction}
S_\text{G}=\int_{\cal M}\td t  \td r ~ \pi_{\ms M} \dot{\ms M} + \pi_{\ms R} \dot{\ms R} - {\ms N_t} {\cal H}_t^{\text{G}} - {\ms N_r} {\cal H}_r^{\text{G}} \,,
\ee
where we used the canonical momenta and defined Hamiltonian densities \cite{Kuchar:1994zk,tbgravity}
\bea\label{momentaaa}
\pi_{\ms M} &=&\frac{\dot{\ms R}-\ms N_r \ms R'}{A \ms N_t}\,,~~~~~~~\hspace{2.5pt}\pi_{\ms R}=\frac{\dot{\ms M}-\ms N_r \ms M'}{A \ms N_t}\,,\\{\cal H}^{\text{G}}_r&=&\pi_{\ms R} \ms R'+\pi_{\ms M} \ms M'\,,~~{\cal H}^{\text{G}}_t=A \pi_{\ms M}\pi_{\ms R}+A^{-1}{\ms M'\ms R'}\,.\nonumber
\eea
The momenta conjugate to the lapse and the shift vanish, enforcing the Hamiltonian constraints
\be
{\cal H}^{\text{G}}_t={\cal H}^{\text{G}}_r=0\,.
\ee
The boundary conditions did not specify any of the unphysical variables ${\ms N_t}$ and ${\ms N_r}$. Correspondingly, the action contains no derivatives of these functions and the coordinate choice is allowed to vary freely everywhere. This has an important consequence: no boundary terms can change the total Hamiltonian or the total energy of the theory, it vanishes as expected in general relativity.

\begin{figure}
\centering
\includegraphics[width=.16\textwidth]{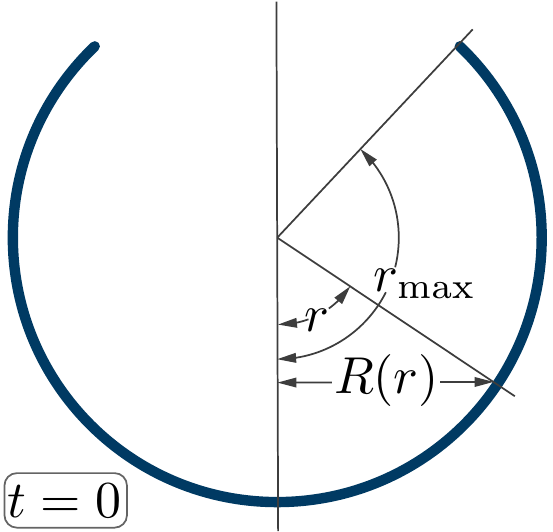}
\caption{\small Closed spatial slice of de Sitter space, a possible initial state of the vacuum decay process.}\label{dsfigure}
\end{figure}
It will be useful to employ coordinates in which the metric component $g_{rr}=1$ is constant, such that the spatial geometry is reflected in the radius ${\ms R}(t,r)$,
\be\label{lambdaeq}
1=g_{rr}=A^{-1}\ms R'^2-A {\pi_{\ms M}^2}\,.
\ee
In this gauge it is straightforward to evaluate the gravitational action along arbitrary trajectories satisfying the constraint equations \cite{tbgravity}. Restricting to solutions of the constraints we have the momenta
\be
\pi_{\ms M}=\eta_\pi A^{-1}{\sqrt{R'^2-A }}\,,~~\pi_{\ms R}=0\,,
\ee
where $\eta_\pi=\pm 1$ denotes the sign of the momentum $\pi_{\ms M}$. Integrating the action (\ref{gravaction}) between initial and final states, we find for classical trajectories \cite{tbgravity}
\bea\label{onshellaction}
S_\text{G}=\int_{r_\text{min}}^{r_\text{max}} \,{\eta_\pi R\over G}\bigg[&&{R'\over 2 }\ln \left({2R'\over A} \left\{R'+\sqrt{R'^2-A}\right\}-1\right)\nonumber\\&&-\sqrt{R'^2-A}\bigg]_{R(t_\text{i},r)}^{R(t_\text{f},r)} \td r\,,
\eea
where $R(t_\text{i,f},r)$ are the initial and final configurations. Following \cite{fmp2} we conveniently illustrate the spatial geometries by plotting the radius $R$ as the distance from the vertical axis, while the radial coordinate $r$ is measured along the curve, see Figure \ref{dsfigure} for a closed slice of de Sitter space.

Clearly, the gravitational action vanishes for any trajectories between classical turning points where the momenta vanish, $\pi_{\ms M}=0$. This differs dramatically from the traditional, non-covariant gravitational action that would assign Euclidean de Sitter space a large action, proportional to its horizon area \cite{tbgravity}. This difference will change the observable vacuum decay rates we find below.

\bigskip
{\it Shell Action}~---~Having obtained the gravitational action that gives rise to a well-posed variational principle for our physical boundary conditions, we now turn to evaluate the action of a domain wall in the general spacetime metric (\ref{metric}). The action for a domain wall with surface energy density $\sigma$ is
\bea\label{ssshell}
S_\text{Shell}&=&-\int_{t_\text{i}}^{t_\text{f}} \td t\, m(\hat{\ms R})\sqrt{g_{rr}(\ms N_t^2-[\dot{\hat{r}}+\ms N_r]^2) }
\\
 &=&\int_{t_\text{i}}^{t_\text{f}} \hat{p}\dot{\hat{r}}-\int_{r_\text{min}}^{r_\text{max}} dr~ ({\ms N_t} {\cal H}_{t}^\text{Shell} +{\ms N_r} {\cal H}_{r}^\text{Shell})~ \td t\,,\nonumber
\eea
where hats denote variables evaluated at the location of the shell \footnote{For example we have $\hat{\ms R}\equiv R(t,\hat{r}(t))$. Some care has to be taken regarding the ordering of hats and derivatives, $\dot{\hat{\ms R}}=\hat{\dot{R}}+\dot{\hat{r}} {\hat{\ms R}}'$.}, and we defined the momentum $\hat{p}$ and energy $\hat{m}=4\pi\sigma \hat{\ms R}$ of the shell, as well as the Hamiltonian contributions ${\cal H}_{t,r}^\text{Shell}$ 
\bea
{\cal H}_{t}^\text{Shell}&=&\sqrt{\hat{p}^2 +\hat{m}^2 \hat{g}_{rr}} \delta(r-\hat{r})\,,~~~{\cal H}_{r}^\text{Shell}=-\hat{p} \delta(r-\hat{r})\nonumber\\
\hat{p}&=& -{\hat{m} (\dot{\hat{r}}+\ms N_r)\sqrt{\hat{g}_{rr}}}/ \sqrt{\ms N_t^2-(\dot{\hat{r}}+\ms N_r)^2}\,.
\eea
The action of the thin shell is identical to the action given in \cite{fmp2, Kraus:1994by, runaways}, but expressed in covariant variables \cite{Kuchar:1994zk,tbgravity}.

\begin{figure}
\centering
\includegraphics[width=.48\textwidth]{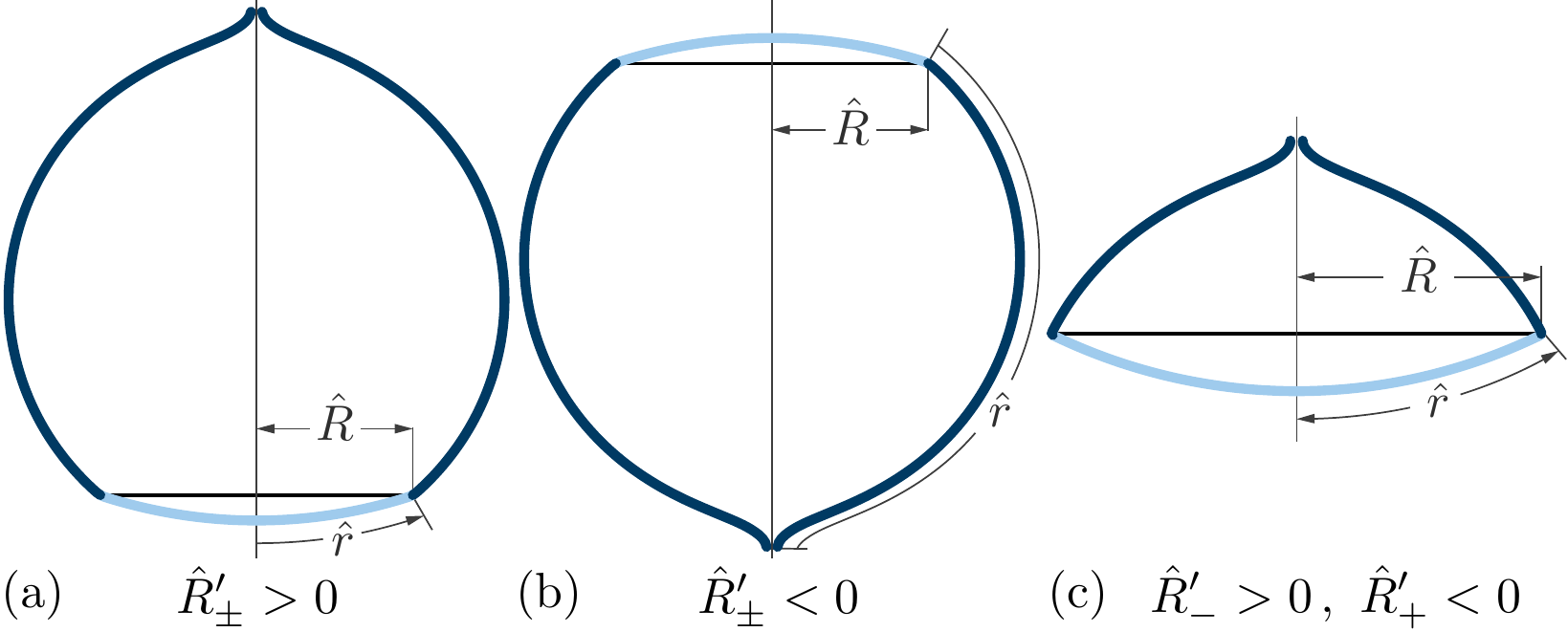}
\caption{\small All possible closed spatial geometries for vacuum bubbles within Schwarzschild-de Sitter region. For $M=0$ these are the final states of the vacuum decay process. (a): true vacuum interior, (b): false vacuum interior, (c): true vacuum interior in strong Gravity regime. Panels (a) and (b) are physically identical and related reversing in/outsides.}\label{bubbles}
\end{figure}
\bigskip
{\it Full Action}~---~In the previous to sections we found the action $S$ for a vacuum bubble in gravity up to the boundary terms $S_\text{B}$
\bea
&&S_\text{G}+S_\text{Shell}=\label{allact}\\
&&\int_{t_\text{i}}^{t_\text{f}} \td t\left[ p\dot{ \hat{r}}+\int_{r_\text{min}}^{r_\text{max}} \hspace{-2pt}\td r\left(\pi_{\ms M} \dot{\ms M} + \pi_{\ms R} \dot{\ms R} - {\ms N_t} {\cal H}_t - {\ms N_r} {\cal H}_r \right)\right]\nonumber
\eea
where we now defined the full Hamiltonian densities as ${\cal H}_{t,r}= {\cal H}_{t,r}^\text{G}+{\cal H}_{t,r}^\text{Shell}$.

In order to explicitly evaluate the action we work in the rest frame of the moving shell (i.e. $\hat{p}=0$ or $\ms N_r=-\dot{\hat{r}}$) and choose coordinates $g_{rr}=-g_{tt}=1$, such that $t$ becomes the proper time. Using (\ref{momentaaa}) and (\ref{lambdaeq}), the velocity in a rest frame traveling along a trajectory $\hat{r}(t)$ is  related to the extrinsic domain wall curvature $\hat{\ms R}'$ as
\be\label{vell}
\dot{\hat{\ms R}}=\hat{\dot{R}}+\dot{\hat{r}} {\hat{\ms R}}'=\sqrt{\hat{\ms R}'^2-\hat{A} }\,.
\ee
The full Hamiltonian constraints ${\cal H}_{t,r}= 0$ become
\bea
0&=&{\cal H}_{t}=A \pi_{\ms M}\pi_{\ms R}+A^{-1}{\ms M'\ms R'}+\hat{m} \delta(r-\hat{r})\,,\nonumber\\
0&=&{\cal H}_{r}=\pi_{\ms R} \ms R'+\pi_{\ms M} \ms M'\,.
\eea
Combining the constraints and integrating across the shell we find the Israel junction condition \cite{Israel:1966rt,bgg,fmp2}
\be\label{junction}
0=\int_{\hat{r}-\epsilon}^{\hat{r}+\epsilon}\td r\left[R''+\kappa R\delta(r-\hat{r}) +\dots \right]=\hat{\ms R}'_+-\hat{\ms R}'_-+\kappa \hat{\ms R}\,,
\ee
where we defined $\kappa\equiv 4\pi G \sigma$, $\hat{\ms R}'_\pm\equiv R'|_{r=\hat{r}\pm\epsilon}$ and the ellipses denote terms that integrate to zero as $\epsilon\rightarrow 0$. We can solve (\ref{vell}) and (\ref{junction}) for the energy conservation equation governing the classical dynamics, as well as the (discontinuous) extrinsic curvatures on either side of the shell
\bea
\ms M_+&=&\ms M_--{4\pi\over 3}\Delta \rho\hat{\ms R}^3+\text{sgn}(\hat{\ms R}'_-)\hat{m} \sqrt{\dot{\hat{\ms R}}^2+\hat{A}_-}- {\kappa^2 \hat{\ms R}^3\over 2G}\,\nonumber \\
\hat{\ms R}'_\pm &=&{\hat{A}_--\hat{A}_+\over2\kappa \hat{\ms R}}\mp {\kappa{\hat{\ms R}}\over 2}\,,\label{energyeqs}
\eea
where $\Delta \rho\equiv \rho_+-\rho_-$ is the positive (negative) vacuum energy density difference for false (true) vacuum decay.  The energy conservation equation deserves some discussion. Each of the terms contributing to the exterior mass $\ms M_+$ has a simple intuitive interpretation. The first term is any contribution from the interior mass, the second term is the contribution from the difference in vacuum energy density, the third term is the rest and kinetic energy contribution and the last term arises from the gravitational self-interaction of the shell. Solving the energy conservation equation yields all possible domain wall trajectories, as well as classical turning points. A detailed discussion of all possible trajectories is given in \cite{bgg,Aguirre:2005xs,Aguirre:2005nt,runaways,BEMtoappear}. Note that the extrinsic curvature ${\ms R}'$ only decreases at the location of the shell. This implies that there are three non-trivial spatial geometries containing vacuum bubbles, illustrated in Figure \ref{bubbles}. 

The extrinsic curvature $ \ms R^\prime$ is discontinuous at the domain wall and changes by a fixed amount (\ref{energyeqs}). In setting up the variational problem for the gravity action $S_\text{G}$, however, we allowed all variables to  vary freely within the region $\cal M$. We have to include the boundary terms $S_\text{B}$ to subtract non-vanishing variations of the gravitational action with respect to $R'$ in order for the Hamilton-Jacobi equation to hold (see also \cite{fmp2, Kraus:1994by} for more details)
\be
-\delta S_\text{B}=\lim_{\epsilon\rightarrow 0}\left({\partial S_{\text{G}}\over \partial \ms R^\prime}\Big|_{r=\hat{r}-\epsilon} - {\partial S_{\text{G}}\over \partial \ms R^\prime}\Big|_{r=\hat{r}+\epsilon}\right) \delta \hat{\ms R}^\prime\,. 
\ee
In our gauge the boundary terms of the classical gravity action (\ref{onshellaction}) become
\bea\label{boundaryterm}
S_\text{B}\hspace{-1pt}=\hspace{-1pt}\int_{\hat{\ms R}(t_\text{i})}^{\hat{\ms R}(t_\text{f})}\hspace{-1pt} {\eta_\pi \hat{\ms R}\over 2G}\bigg[&&\ln\Big({2\hat{A}_+^{-1}\hat{\ms R}'_+\Big\{\hat{\ms R}'_++\sqrt{\hat{\ms R}^{\prime2}_+-\hat{A}_+} \Big\}}-1\Big)\nonumber\\&&-\Big(\hat{A}_+\rightarrow \hat{A}_-\,,~\hat{\ms R}'_+\rightarrow \hat{\ms R}'_-\Big)\bigg]\td \hat{\ms R}\,.
\eea
Recall from above that the gravitational action $S_\text{G}$  vanishes for all solutions between stationary points. Furthermore, the shell action $S_\text{Shell}$ in (\ref{ssshell}) vanishes in the rest-frame of the domain wall, where $\hat{p}=0$. In our gauge choice, this leaves the boundary term (\ref{boundaryterm}) as the only non-zero contribution to the action for trajectories between turning points,
\be\label{fullaction}
S=S_\text{B}\,.
\ee
As a consistency check, we easily recover the energy conservation equation (\ref{energyeqs}) by varying the action $S_\text{B}$.

\bigskip
{\it Classical Dynamics}~---~Before moving on to the discussion of vacuum decay, we briefly review the classical dynamics of true- and false-vacuum bubbles in (anti) de Sitter spacetimes, i.e. $\ms M_\pm=0$. References \cite{Aguirre:2005xs,Aguirre:2005nt,runaways,bgg,BEMtoappear} provide a detailed discussion of the general dynamics and causal structure.

In the massless limit the  constraints (\ref{energyeqs}) for the domain wall become
\be
\hat{\ms R}'_\pm = {\hat{\ms R}\over 6\kappa} \left[8\pi G{\Delta \rho}\mp {3\kappa^2} \right], ~
\hat{\ms R}^4 \dot{\hat{\ms R}}^2=\hat{\ms R}^4\left({\hat{\ms R}^2\over \hat{\ms R}^2_2}-1\right),\label{energyeqsmassless}
\ee
where the shell velocity vanishes at the turning points
\be\label{radius}
\hat{\ms R}_1=0\,,\hat{\ms R}_2=\frac{6 \kappa}{\sqrt{(8 \pi G \Delta \rho)^2+48 \pi  G \kappa^2 (\rho_-+\rho_+)+9 \kappa^4}}.
\ee
A domain wall at the inner turning point, $\hat{\ms R}_1=0$ corresponds to the initial, metastable configuration of vacuum decay. In order for a finite-size bubble to form, a tunneling event through the classically forbidden region $\hat{\ms R}_1<\hat{\ms R}<\hat{\ms R}_2$ has to occur that leads to a classically growing bubble on the unbound trajectory
\be\label{unbound}
\hat{\ms R}(t)=\hat{\ms R}_2 \cosh(t/\hat{\ms R}_2)\,.
\ee

The sign of the extrinsic curvature, $\hat{\ms R}'_\pm$, indicates whether the domain wall curves towards the bubble interior or exterior as seen from either side of the wall. A true vacuum bubble $\rho_-<\rho_+$ observed from the interior always curves towards the interior  (i.e. $\hat{\ms R}'_->0$, panel (a) in Figure \ref{bubbles}). When observed from the outside, however, the true vacuum bubble can curve towards the interior in the weak-gravity limit (i.e $\hat{\ms R}'_+>0$  when $\Delta \rho>6 \pi G \sigma^2$, panel (a) in Figure \ref{bubbles}), or towards the exterior in the strong-gravity limit (i.e $\hat{\ms R}'_+<0$ when $\Delta \rho<6 \pi G \sigma^2$,  panel (c) in Figure \ref{bubbles}). The definition of what we call interior and exterior is clearly arbitrary, so an equivalent statement applies for false vacuum bubbles (in the weak gravity limit these bubbles have $\hat{\ms R}'_\pm<0$, see panel (b) in Figure \ref{bubbles}). Some classically allowed and forbidden domain wall trajectories will cross spacetime horizons. For example, a classically expanding true vacuum bubble will cross the cosmological horizon after some finite proper time has elapsed at the wall.

The classically growing domain wall trajectory  (\ref{unbound}) does not always exist. When both vacuum energy densities are non-positive, it is possible for the classical turning point radius $\hat{\ms R}_2$ to diverge. When this happens, only the bound solution at $\hat{\ms R}=\hat{\ms R}_1=0$ exists, and no  vacuum decay can occur: gravity has stabilized the vacuum \cite{Coleman:1980aw}.

\bigskip
{\it Vacuum Decay Rates}~---~In the previous section we found two classical turning points of the domain wall trajectory: the vacuum configuration $\hat{\ms R}=0$ in which there is no bubble, and the bubble configuration $\hat{\ms R}=\hat{\ms R}_2$ in which a bubble begins to classically expand indefinitely. The vacuum decay rate is proportional to the transition probability  for the bubble to tunnel through the classically forbidden region rather than being reflected. In the semiclassical approximation we have the decay rate (\ref{tunnelingpintro}), where the tunneling exponent is the Euclidean bounce action \cite{Coleman:1977py,Coleman:1980aw}
\be\label{tunnelingexp}
B=-2i \int_{{\hat{\ms R}=\hat{\ms R}_1}}^{{\hat{\ms R}=\hat{\ms R}_2}}  dS\,.
\ee
For vanishing masses, $\ms M_\pm=0$, we can analytically evaluate the tunneling exponent. Using (\ref{energyeqs}-\ref{fullaction}) we find
\bea\label{beq}
 &&B=\\&&{3\eta_\pi\over 16 G^2}\hspace{-1pt}\bigg[\hspace{-1pt} \hat{\ms R}_2\hspace{-1pt} \frac{ \Delta \rho^2\hspace{-1pt}+\hspace{-1pt}6 \pi  G \sigma ^2 (\rho_-\hspace{-1pt}+\hspace{-1pt}\rho_+)}{3 \rho_- \rho_+ \sigma }\hspace{-1pt}+\hspace{-1pt}\frac{  \text{sgn}(\hat{\ms R}^\prime_{+})}{ \rho_+}\hspace{-1pt}-\hspace{-1pt}\frac{   \text{sgn}(\hat{\ms R}^\prime_{-})}{\rho_-}\hspace{-1pt}\bigg]\hspace{-.5pt},\nonumber
\eea
where the extrinsic curvatures are evaluated at $\hat{\ms R}_2$ and their signs are  given by
\be
 \text{sgn}(\hat{\ms R}^\prime_{\pm})=\text{sgn}(\Delta \rho\mp6 \pi  G \sigma ^2)\,.
\ee
The vacuum decay rate (\ref{beq}) is the main result of this paper and has not previously appeared in the literature. 

We have not yet determined the sign $\eta_\pi$ appearing in the tunneling action. Since we are only considering the action at vanishing mass, which does not allow for time evolution, it is impossible to determine the sign by demanding an out-going wavefunction. However, on physical grounds we assume that the tunneling probability is small, giving
\be\label{signb}
\text{sgn}(B)\equiv 1= \text{sgn}(\eta_\pi\hat{\ms R}^\prime_{+}\hat{\ms R}^\prime_{-})\,\rightarrow \eta_\pi= \text{sgn}(\hat{\ms R}^\prime_{+}\hat{\ms R}^\prime_{-})\,.
\ee
We now discuss some interesting features of the decay rate. For weak-gravity false vacuum decays, where $\text{sgn}(\hat{\ms R}^\prime_{\pm})>0$, the tunneling exponent (\ref{beq}) precisely reproduces the finding of Coleman and de Luccia \cite{Coleman:1980aw}, but our result applies more generally. 

In the $G\rightarrow 0$ limit the tunneling exponent for both true- and false-vacuum decay becomes
\be\label{weakg}
B=\frac{27 \pi ^2  \sigma^4}{2 \left|\Delta \rho\right| ^3}+{\mathcal O}(G)\,.
\ee
Without gravity true vacuum decay is not a possible solution, so there is no non-gravitational result we could compare our rate to. In general relativity, however, true vacuum decay is possible and its rate is identical to the false vacuum decay rate,
\be
\Gamma_{\rho_+>\rho_-}=\Gamma_{\rho_+<\rho_-}\,,
\ee
showing bubbles of de Sitter space can nucleate in Minkowski space. 

Our result differs from the literature \cite{Coleman:1980aw,Brown:1987dd,Brown:1988kg,Lee:1987qc,fgg} because we are using different boundary conditions. Instead of holding fixed the unobservable global three-geometry, which assigns a slice of de Sitter space a large action, we demand that the observable covariant mass $\ms M$ vanishes on either side of the shell, which assigns the vacuum a vanishing action.

\bigskip
{\it Negative Mode}~---~It is generally believed that for the bounce to correspond to barrier penetration, the second variation of the Euclidean action has one and only one negative eigenvalue. A rough argument is that the negative eigenvalue is necessary for the non-perturbative contribution to the vacuum energy to be imaginary, which destabilizes the vacuum \cite{Coleman:1987rm}. It has been argued that no such negative mode exists for certain parameter regimes of vacuum decay in gravitational theories that hold fixed the induced boundary metric \cite{Marvel:2007pr,Weinberg:2012pjx,Yang:2012cu}.

Finding all eigenvalues of the second variation is hard. Instead, we merely evaluate the second derivative of a set of trajectories parametrized by the maximum radius $\hat{\ms R}(t_\text{f})$, where $\hat{\ms R}(t_\text{f})=\hat{\ms R}_2$ corresponds to the bounce solution. Expanding around the bounce and using (\ref{boundaryterm}) and (\ref{signb}) yields
\be
{\partial^2 B\over \partial \hat{\ms R}(t_\text{f})^2}\propto-{\eta_\pi\over\hat{\ms R}^\prime_{+}\hat{\ms R}^\prime_{-}} <0\,,
\ee
This shows that the second variation of the Euclidean action always has at least one negative eigenvalue.

\bigskip
{\it Acknowledgments}~---~We are particularly thankful to Kate Eckerle and Ruben Monten for many useful discussions and collaboration on a related work \cite{BEMtoappear}. We thank Frederik Denef, Thomas Hartmann, Austin Joyce, Liam McAllister, Henry Tye, Oliver Janssen, Matthew Kleban and Erick Weinberg for useful discussions. This work was supported in part by DOE under grants no. DE-SC0011941 and DE-SC0009919 and by the Simons Foundation SFARI 560536.

\bibliographystyle{klebphys2}
\bibliography{bubblerefs}

\end{document}